\RequirePackage{fix-cm}
\documentclass[smallextended]{svjour3}       
\smartqed  
\usepackage{graphicx}
\def\msun{M$_\odot$}
\def\iso#1{$^{#1}$}

\usepackage[numbers]{natbib}
\usepackage{url} 
\usepackage{hyperref}
\usepackage[dvipsnames]{xcolor}
\usepackage[normalem]{ulem}

\newcommand{\aap}{{\it A\&A}}
\newcommand{\aapr}{{\it A\&A Rev.}}
\newcommand{\apj}{{\it ApJ}}
\newcommand{\apjl}{{\it ApJL}}
\newcommand{\apjs}{{\it ApJ Suppl. Ser.}}
\newcommand{\mnras}{{\it MNRAS}}

\newcommand{\nat}{{\it Nature}}

\newcommand{\araa}{{\it Ann. Rev. Astron. Astrophys.}}
\newcommand{\prl}{{\it Phys. Rev. Letters}}
\newcommand{\prc}{{\it Phys. Rev. C}}
\newcommand{\pasa}{{\it Publ. Astron. Soc. Austr.}}
\newcommand{\gca}{{\it Geochim. Cosmochim. Acta}}

\newcommand{\ssr}{{\it Space Sci. Rev.}}

\begin{document}

\authorrunning{Lugaro et al.}
\titlerunning{Isotopic variations of zirconium and neodymium}

\title{Predictions from $s$-process AGB models of the isotopic variations of zirconium and neodymium for comparison to 
bulk meteorites}




\author{Maria Lugaro$^{1,2,3,4}$, Giulia C. Cinquegrana$^{4}$, Bal\'azs Sz\'anyi$^{1,2,5}$, James M. Ball$^{6}$, Borb\'ala Cseh$^{1,2}$, Mattias Ek$^{6}$, Amanda I. Karakas$^{4}$, Maria Sch\"onb\"achler$^{6}$, and John C. Lattanzio$^{4}$}



\institute{
$^{1}$Konkoly Observatory, HUN-REN Research Centre for Astronomy and Earth Sciences, H-1121 Budapest, Konkoly Thege M. \'ut 15-17, Hungary \\ 
           $^{2}$CSFK, MTA Centre of Excellence, Budapest, Konkoly Thege Miklós út 15-17, H-1121, Hungary \\ 
           $^{3}$ELTE E\"{o}tv\"{o}s Lor\'and University, Institute of Physics, Budapest 1117, P\'azm\'any P\'eter s\'et\'any 1/A, Hungary\\
          $^{4}$School of Physics and Astronomy, Monash University, VIC 3800, Australia\\ 
           $^{5}$Department of Experimental Physics, Institute of Physics, University of Szeged, H$-$6720 Szeged, Dóm tér 9, Hungary \\ 
           $^{6}$Department of Earth and Planetary Sciences, Institute of Geochemistry and Petrology, ETH Zürich, Zürich, Switzerland\\ 
}

\date{Received: date / Accepted: date}

\maketitle

\begin{abstract}
Bulk meteoritic data show isotopic variability of $slow$-neutron-capture ($s$-process) origin in several elements heavier than Fe. One peculiar feature is that the lighter $s$-process elements (e.g., Zr and Mo) present larger anomalies than the heavier $s$-process elements (e.g., Nd and W). To address this observation, we compared Zr and Nd data to model predictions of the $s$-process abundances at the surface of low-mass asymptotic giant branch (AGB) stars of initial metallicity from solar to twice solar. We found that the relative magnitude of the isotopic variability between these two elements can be matched by models of AGB stars of super-solar metallicity. The match is favoured by stronger convective overshoot, leading to a deeper dredge-up of the H-rich envelope into the He-rich region, and/or a smaller ($\sim$half than standard) mass of the region rich in the \iso{13}C nuclei that produce free neutrons via the \iso{13}C($\alpha$,n)\iso{16}O reaction. We conclude that nucleosynthesis in AGB stars can match the difference in the magnitude of the bulk meteoritic variations in Zr and Nd, provided that super-solar metallicity stars are the original site of these signatures. 
The AGB stars that produced such variations could have belonged to the current population of old, super-solar metallicity stars seen in the galactic solar neighbourhood. 
\end{abstract}
%
%
\section{Introduction}
\label{sec:intro}
In the past couple of decades it has been possible to analyse meteoritic materials to such a high level of precision that isotopic variations of stellar nucleosynthetic origin have been identified not only in the largest and oldest inclusions (the calcium-aluminium-rich inclusions, CAIs), but also in bulk rock samples of different meteorites and, recently, even in smaller inclusions such as chrondrules, and in the matrix \citep[see reviews, e.g.,][]{tissot25rev,Schonbachler25REV}. 
This isotopic variability resulted from heterogenous distribution of a small amount of stellar material (probably stardust) in the protoplanetary disc and provides us with an identifiable signal that can be followed to investigate the molecular cloud where the Sun was born, the Sun's accretion history, the evolution of the protoplanetary disc, and the formation of the planets  \citep[see reviews, e.g.,][]{REVqin16,REVdauphas16,REVbermingham20,REVmezger20,REVkleine20,REVbizzarro25}. One of the first and most obvious nucleosynthetic isotopic signatures in meteoritic bulk rocks is represented by isotopic variations produced by the $slow$ neutron-capture process (the $s$ process) in asymptotic giant branch (AGB) stars \citep[e.g.,][]{dauphas02,dauphas04,ek20}. 

In spite of the large number of data available for several elements heavier than Fe affected by the $s$ process, a full, detailed comparison between meteoritic data and AGB models is still lacking. Most of the data interpretation is based on results from models of AGB stars and galactic chemical evolution aimed at predicting the solar $s$-process distribution \citep{bisterzo11PaperII,bisterzo14}. This is inaccurate, because the AGB stars that produced the material that built up the $s$-process solar abundance distribution over the entire lifetime of the Galaxy, are not the same AGB stars that contributed individual stardust grains to the early Solar System, whose heterogenous distribution is responsible for the observed isotopic variability. This is demonstrated by the fact that stardust silicon carbide (SiC) grains from AGB stars recovered from meteorities do not carry solar isotopic ratios for the isotopes that are exclusively, or almost exclusively, produced by the $s$ process, such as \iso{134}Ba/\iso{136}Ba and \iso{88}Sr/\iso{86}Sr \citep[e.g.,][]{liu14Ba}. 

Moreover, AGB stars of around solar metallicity are the favoured parent stars of most meteoritic stardust because interstellar dust is predicted to survive typically only a few hundred Ma in the interstellar medium. Theoretical lifetime estimates are between 30 Ma to 1 Ga, when considering current uncertainties \citep{jonesnuth11}, in agreement with observational data from relatively large $>$1 $\mu$m stardust grains, showing that very few grains have ages longer than 1 Ga \citep[e.g.,][]{heck20}. 
In any case, here we also present some low-metallicity stellar models to check if their predictions may be in agreement with our target observations.

Roberto Gallino was a pioneer 
in using detailed AGB models to interpret meteoritic data, not only in the form of stardust grains \citep[e.g.,][]{gallino90,gallino97,lugaro99,lugaro03grains,nittler08,liu22}, but also from leachates \citep[meteoritic residuals from acid treatments][]{reisberg09leachatesOs} and bulk rock analysis \citep{qin08bulkW,akram15bulkZr}. Following this line of research, we have initiated a long-term project aimed at comparing meteoritic data to detailed $s$-process models, based on the method described in \cite{lugaro23}. This kind of detailed comparison will benefit both the use of meteoritic data to investigate the formation of the Solar System, and the understanding of AGB stars and the $s$-process within them. 

In this paper, 
we present the first interpretation of the observed Zr and Nd relative variability in terms of $s$-process variability carried by AGB SiC stardust. This hypothesis is based on the fact that both the Zr and the Nd variability can be explained by the $s$ process and that $s$-process variability is seen in several other elements in meteoritic sample, including acid leachates \citep[e.g.][]{dauphas04cosmicMoRu,burkhardt12LeachatesBulkMoW,fischergodde15bulkleachatesRu,worsham19IronRuMoW,elfers20bulkleachatesZrHf}. This hypothesis is also supported by the existence in the meteoritic inventory of stardust SiC grains showing the $s$-process signature in many elements including Zr and Nd, i.e., depletions in \iso{96}Zr and \iso{148}Nd \cite{richter92, nicolussi98,liu14Zr}.
We follow-up on the hypothesis that AGB stars of metallicity higher than solar contributed significantly to the $s$-process variability in bulk meteorities \cite{ek20}. Some of us have already showed that predictions from super-solar-metallicity AGB stars best fit stardust SiC single grain data for elements such as Sr, Zr, and Ba \citep{lugaro18grains,szanyi25}, at the same time providing a self-consistent interpretation of the abundance trends with metallicity observed in Ba stars, which accreted $s$-process elements from a former AGB companion \citep{lugaro20}.  

The key here is the comparison of the relative variations observed in bulk meteorites for two elements belonging to the first and to the second $s$-process peaks: Zr (close to the neutron magic number 50) and Nd (close to the neutron magic number 82), respectively. If the origin of their relative variations can be explained by the $s$ process in AGB stars, there is no need to invoke different dust carriers and their chemical processing in the protoplanetary disc and in the interstellar medium. Moreover, if confirmed to have had metallicity higher than solar, these AGB stars could have belonged to the population of old, high-metallicity stars currently observed in the galactic solar neighbourhood \cite{nissen20}, providing observational constraints on its history.

In the following introductory subsections we describe the process of neutron captures in AGB stars (\ref{sec:agb}) and the meteoritic data of interest (\ref{sec:data}). We describe our methods to calculate the AGB surface abundances and the 79 stellar models used for the present analysis in Section~\ref{sec:codes}, and how the AGB surface abundances are post-processed to derive the quantities needed for the data comparison in Section~\ref{sec:norm}. In Section~\ref{sec:results} we present and discuss the results, and in Section~\ref{sec:conclusions} we summarise our conclusions and propose future work directions.  

\subsection{The $s$ process in AGB stars}
\label{sec:agb}
The AGB phase of stellar evolution is reached by low and intermediate-mass stars ranging in mass from roughly 0.8 to 8 \msun, depending on
the initial composition, after evolution through core hydrogen and helium burning \citep[e.g.,][]{busso99,herwig05,karakas14dawes}. Nucleosynthesis during the AGB phase is mostly driven by He-shell instabilities, or thermal pulses, which can be followed by episodes of mixing between the He-rich, H-exhausted core and the H-rich envelope. This mixing is known as the `third dredge-up' and is responsible for enriching the stellar surface with carbon and elements heavier than Fe produced by the $s$ process, in the presence of free neutrons \cite{REVLugaro23}. 

The main way to produce neutrons in AGB stars is through $\alpha$-capture on $^{22}$Ne and $^{13}$C in the He-rich region. For the low-mass AGB stars (less than $\sim$3 \msun) of metallicity around solar, the maximum temperature is below $\sim 3 \times 10^8$ K, barely enough to activate the $^{22}$Ne source. Instead, it is the $^{13}$C($\alpha$,n)$^{16}$O reaction that provides the neutrons. This, in turn, requires the presence of $^{13}$C in significant amounts. This isotope can be produced by proton capture on the abundant $^{12}$C in the He-rich region, via the $^{12}$C(p,$\gamma)^{13}$N($\beta^+ \nu)^{13}$C reaction chain. The challenge for the star is to bring protons and $^{12}$C together in the appropriate amounts. This is believed to occur during the third dredge-up phase of thermal pulses on the AGB. Here, we have a H-rich envelope extending inwards, into the He-rich region enriched in $^{12}$C produced by partial He burning in the previous thermal pulses. The most likely location for this mixing process is the interface between the H-rich convective envelope and the $^{12}$C-enriched radiative region. Notably, these two must mix only {\it partially\/}, and not homogeneously because if too many protons are present, then the CN cycle produces $^{14}$N rather than $^{13}$C (by one more proton capture on the $^{13}$C), so it is necessary that only small numbers of protons are mixed into the $^{12}$C-enriched region. Various mechanisms have been suggested for this partial mixing, such as semiconvection, gravity waves, magnetic buoyancy, and various instabilities, which can be expected at the borders of convective regions \cite[as summarised in the reviews by][]{herwig05,karakas14dawes,REVLugaro23,dominguez26}. It is fair to say that none of these can be modelled accurately at present, and one usually resorts to some imposed mixing process. The $^{13}$C($\alpha$,n)$^{16}$O reaction is subsequently activated at temperatures above $\sim$0.9 MK, usually in radiative conditions before the onset of the following thermal pulse. In some cases, however, the temperature is not high enough to burn all the $^{13}$C nuclei. Some of them will be ingested and release neutrons inside the convective thermal pulse region, as we will see occurring in some of our models \cite[and as also reported in some FRUITY models][]{cristallo11FRUITY}.

In this paper, we will consider models of low-mass AGB stars, which we define as those with initial masses up to $\sim4$ \msun. These stars make up the majority of the AGB stars that we observe in galaxies, owing to the initial mass function favouring low-mass stars, and also their relatively long lifetime on the thermally pulsing AGB phase ($\sim2$ Ma). The surface composition of low-mass AGB stars is characterised by enrichments in C (in the form of $^{12}$C), which lead to an increased C/O ratio, along with N, F, and $s$-process elements, as observed \citep{wallerstein98,abia02}. 
Low-mass AGB stars with third dredge-up can become carbon rich, i.e., the number density of carbon atoms exceeds the number density of oxygen atoms (C/O $\ge 1$). Observations of carbon rich AGB stars show them to be bright, infrared objects often losing mass through dusty outflows \citep{hofner18}.

Intermediate-mass AGB stars with initial masses between roughly 4 \msun\ to 8 \msun\ are considerably rarer in AGB populations \citep{wood83}. These stars are predicted to experience a second dredge-up during the early AGB phase, before the thermal pulses begin, which is responsible for significant enrichments in surface He, N, and Na. Differently to the lower-mass AGB stars, intermediate-mass AGB stars also experience proton-capture nucleosynthesis at the base of the convective envelope, known as hot bottom burning \citep{karakas03AlMg,ventura13}. The surface chemistry of intermediate-mass stars is therefore characterised not only by the third dredge-up, but also by proton-capture nucleosynthesis, which results in C/O $< 1$ as a result of CN cycling burning C into N, with perhaps some addition of material from the He-shell, noting that the envelope mass is larger and the efficiency of TDU is lower than in their lower-mass counterparts. 
Similar surface chemistry and nucleosynthesis is also predicted for super-AGB stars, which are stars with initial masses between 8 \msun\ to 10 \msun, which experience core carbon burning before ascending the AGB \citep{doherty14a,doherty15}. The final fate of super-AGB stars is likely an O$-$Ne white dwarf but their fate may also be electron-capture supernovae, for the most massive stars in that range \citep{doherty15}. In contrast the fate of stars with initial masses up to about 8 \msun\ is to become a C-O core white dwarf. We do not consider intermediate-mass AGB and super-AGB stars here as, on top of producing mostly O-rich dust, they are not believed to have contributed significantly to the $s$ process in the Galaxy \citep[see, e.g., Table 1 of][]{bisterzo14}. Mostly due to their small He-rich regions and very large envelopes, these stars do not generally produce high yields of $s$-process elements \cite{cristallo15FRUITYHighM,karakas16}.

\subsection{Current constraints from the nucleosynthetic isotopic composition of bulk meteorites}
\label{sec:data}

The meteoritic data we consider here to derive the observational constraint to be matched by the models is from a type of meteorites called chondrites. These rocks are agglomerations of dust from the early Solar System, which remained relatively pristine with varying degrees of thermal and aqueous alteration from $\sim$4.6 Ga until the present day \citep{Scott14chapter}. These meteorites contain presolar stardust grains inherited from stars that contributed dust to proto-solar nebula \citep{liu25}, and refractory inclusions, such as CAIs, which represent some of the earliest solid matter to have condensed within the proto-planetary disc \cite[e.g.,][]{MacPherson14chapterCAI}. Isotopic variations in presolar stardust grains, in particular, are extremely anomalous when compared to the bulk Solar System and their presence in the rocks affect the bulk (i.e., average) composition of chondritic meteorites.

In bulk chondrites, the lighter (below Ba) refractory elements heavier than Fe show larger $s$-process isotope variations than their heavier counterparts (see, e.g., Figure 9 of \citep{elfers20bulkleachatesZrHf}, Figure 4 of \citep{burkhardt12LeachatesBulkMoW}, and \cite{ball24}). These lighter and heavier elements vary together with the abundances of the elements belonging to the first and second $s$-process peaks, at neutrons magic numbers 50 and 82, respectively. This behaviour indicates either (i) that these elements were produced by different stellar sources and carried by different types of dust, which behaved differently in the protosolar disc, 
or (ii) that the AGB stars that contributed 
presolar stardust to the molecular cloud from which the Solar System formed preferentially produced the lighter rather than the heavier $s$-process elements. The latter hypothesis is suggested to be related to the metallicity of the AGB star \citep{ek20}. The \iso{13}C neutron source forms via proton captures on the \iso{12}C produced by He burning (Section~\ref{sec:agb}), therefore, it is `primary' in nature, i.e., it depends only on the initial H and He content of the star, and not on the stellar metallicity. The initial abundance of Fe seed nuclei, instead, is of course metallicity dependent. How many neutrons are captured by these Fe seeds controls the number of free neutrons and the efficiency of the $s$-process in reaching the heavier elements \citep{clayton88,gallino98,cseh18}. To explore this hypothesis further we compare AGB models of metallicity from solar to twice solar to the $s$-process isotopic variability observed in a light (Zr) and a heavy (Nd) $s$-process element. Through this we aim to add constraints on the origin of such variability and on the AGB stars that contributed stardust to the galactic solar neighbourhood around the time of the Sun's birth.

In cosmochemistry, the 50\% condensation temperature (50\% T\textsubscript{c}), i.e., the temperature at which 50\% of each element has condensed into solids from a cooling gas of solar composition, is used as a proxy for volatility, i.e., the 
tendency to vaporise at low temperatures \citep{lodders25volat}. The more volatile an element is, the lower its 50\% T\textsubscript{c}, and  
the more its nucleosynthetic variations may have been overprinted by thermal and chemical processes \cite{toth20bulkleachatesCd}. For this reason, here we focus on two refractory elements \citep[with 50\% T\textsubscript{c} $\ge$ 1300 K][]{lodders03}, which are more likely to have retained their nucleosynthetic signatures unperturbed by disc processes. The elements Zr and Nd are the perfect candidate elements to be chosen for comparison with models of AGB stars because not only they have 50\% T\textsubscript{c} = 1741 K and 1602 K, respectively, \citep{lodders03}, but also they have been measured to high precision in bulk chondrites. Moreover, leachate experiments have been conducted on chondrites for both Zr \citep{schonbachler05leachatesZr,bizzarro25ice} and Nd \citep{qin11crsrbasmnd} and the Zr data have been compared with available in-situ studies of large presolar SiC from AGB stars \citep{nicolussi98}. The results clearly demonstrate that presolar SiC from AGB stars is the primary (but by no means only) carrier of isotopic variation in bulk chondrites for these two elements. 

\section{Methods and Models}
\label{sec:methodology}

\subsection{The Monash codes and the nuclear network}
\label{sec:codes}

The structure of the stellar models was computed with the Monash stellar evolution code, adapted from the Mt. Stromlo code \citep{LattanzioThesis, Lattanzio86, Frost96, Karakas07}\footnote{A detailed discussion of the numerics has been recently provided in \citep{cinquegrana2022bridging}.}. The full model grid and associated nucleosynthesis yields will be presented separately (Cinquegrana, Sz\'anyi, Lugaro, \& Karakas, in preparation). Each model is evolved from the zero-age main sequence to the end of the thermally pulsing AGB, or until convergence difficulties prevent further evolution. 
These difficulties may arise during the superwind phase, i.e., when the rest of the envelope is stripped very fast (with mass loss rate of the order of $10^{-5}$ to $10^{-4}$ \msun/yr), according to the mass loss prescription that we use \citep{vw93}.
Three initial metallicities ($Z$) are considered in this study $Z=0.014, 0.02$ and $0.03$, with initial He abundances defined by:
\begin{equation} \label{delYdelZ}
Y_\mathrm{i} = Y_0 + \frac{\Delta Y}{\Delta Z},Z_\mathrm{i}.
\end{equation}
We take a primordial He abundance of $Y_0=0.2485$ \citep{Aver13} and He-to-metal enrichment ratio of $\rm \Delta Y/\Delta Z=2.1$ \citep{Casagrande07}. We model mass loss on the red giant branch using the Reimer \citep{R75}'s prescription with an efficiency of $\eta=0.477$ \citep{Mcdonald15mass}. On the AGB, we use the empirical prescription of \citep{Vassiliadis94}. Convection is approximated with the Mixing-Length Theory \citep{Prandtl25, Vitense53} with a mixing-length parameter $\alpha=1.86$ and assuming instantaneous mixing in convective regions. Low-temperature opacities are the same as we computed for \citep{karakas2022most}, calculated using the {\AE}SOPUS tool \citep{marigo2009low}. These opacities include CNO variations relevant to the thermally pulsing AGB evolution. High-temperature opacities are from OPAL \citep{Iglesias96} adopting the solar mixture from \citep{lodders03}. 

We define convective boundaries using the convective-neutrality algorithm introduced in \citep{Lattanzio86}. Previous works \cite{lattanzio89,karakas16} have shown that this algorithm results in the activation of the third dredge-up, however, it underpredicts the efficiency of third dredge-up mixing in AGB stars of mass $\sim$1.5-2 \msun\ when compared, for example, to the carbon star luminosity function of the Galaxy and the Magellanic Clouds \citep{kamath2012evolution,osborn25}. To probe the implications of increasing the efficiency of the third dredge-up, and of  
lowering the core mass at which it first appears, we also computed models with a 
parametrised convective overshoot, where we extend the base of the convective envelope during the third dredge-up downward by a fixed fraction ($N_\mathrm{ov}$) of the pressure scale height:
\begin{equation}
    H_{\rm p} = -P \frac{d r}{d P}.
\end{equation}
Implementing convective overshoot of up to $N_\mathrm{ov} \approx 3\, H_\mathrm{p}$ at the base of the convective envelope during the third dredge-up was required to reproduce the observed O- to C-rich transition luminosities in stellar clusters in the Magellanic Clouds \citep{kamath2012evolution}.


The nucleosynthesis models were calculated using the Monash post-processing nucleosynthesis code \citep{ca93}, which calculates abundance changes due to time-dependent mixing\footnote{This employs an advective mixing scheme based on the mixing velocities calculated using the mixing length framework in the structure code. In any case, these velocities are high enough to ensure that the envelope is well mixed within the timescale of weeks, so in relation to the $s$-process elements it is effectively as if we were using almost instantaneous mixing.} and nuclear burning on the basis of stellar structure input (temperature, densities, and mixing velocities) calculated as described above. The nuclear reaction network includes 328 isotopes and 2,351 reactions and it is similar to Set 2 of \citep{sz25}. In the context of the isotopes discussed here, the (n,$\gamma$) reaction rates of Zr isotopes are from \citep{lugaro14Zr}, while the rates of Nd isotopes are the \textit{ka02} labelled fits from JINA reaclib database \citep{cy10}. The $\mathrm{\beta^-}$-decay rates of unstable Zr and Nd isotopes are from the NETGEN \citep[nuclear NETwork GENerator,][]{jo01,ai05,xu13} compilation, specifically, the rates of \iso{93}Zr and \iso{147}Nd are based on the work of \citep{goriely99} and the rate of \iso{95}Zr is based on the work of \citep{takahashi87}. The neutron-source reactions $\mathrm{^{13}C(\alpha,n)^{16}O}$ and $\mathrm{^{22}Ne(\alpha,n)^{25}Mg}$ are the JINA reaclib fit of the rate by \citep{heil08}, which is close to the more recent values from the LUNA and JUNA underground experiments \citep{ciani21LUNAc13an,gao22JUNAc13an}; and the rate by \citep{iliadis10}, respectively. The $\mathrm{^{22}Ne(\alpha,n)^{25}Mg}$ rate is not yet available from underground experiments. The \textit{s}-process enhancements observed in AGB stars, their companions, and their progeny require the formation of a partial mixing zone at the top of the He-rich region, where protons from the convective envelope are captured by the abundant $\mathrm{^{12}C}$, resulting in a $\mathrm{^{13}C}$ `pocket' \citep[see Section~\ref{sec:agb} and e.g.,][]{st97, ga98, go00, lugaro03sprocess}. 
We include this required partial mixing in the nucleosynthesis code following 
a method described in detail by \citep{karakas16,buntain17}. In summary, 
at the end of each third dredge-up episode, 
the proton abundance is set to decrease exponentially with mass depth, from the envelope H abundance of $\sim$ 0.7, down to 0.00007 at a given location in the He-rich region. The difference in mass between this location and the base of the convective envlope represents the mass extention of the partial mixing zone, which we treat as a free parameter called $M_\mathrm{mixed}$.

\subsection{The stellar models}
\label{sec:models}



Tables \ref{tab:models0p014}, \ref{tab:models0p02}, \ref{tab:models0p03} list the 80 models considered here. Four parameters characterise each model, three of them are set in the evolutionary calculations: the metallicity ($Z$), which we set to 0.0028, 0.007, 0.014\footnote{This is the solar metallicity assumed in our models \citep{asplund09}, which is roughly 30\% lower than the latest reported value 0.0187 \cite{lodders25}, due to the latest higher C and O abundances.}, 0.02, 0.03, 0.04, and 0.05, the initial stellar mass ($M$), which we vary from 1.75 to 4 $\mathrm{M}_\odot$, 
and the extent of forced convective overshoot $N_\mathrm{ov}$ during third dredge-up episodes, which we vary between 0 and 4, in units of the pressure scale height $H_\mathrm{p}$. 
The mass of mixed zone $M_{\rm mixed}$ leading to the formation of the \iso{13}C pocket, instead, is set in the post-processing nucleosynthesis calculations and varied between $5 \times 10^{-4}$ and $2 \times 10^{-3}$ \msun. This means that each stellar structure model of a given same mass, metallicity and $N_\mathrm{ov}$ was run several times through the nucleosynthesis post-processing code with different values of $M_{\rm mixed}$. As mentioned at the end of Section~\ref{sec:agb}, we focus on models of AGB stars in the mass range that is predicted to become C-rich and enriched in the $s$-process elements \citep[see discussion, e.g., in][]{karakas16,cseh22}. As mentioned in Section~\ref{sec:data} we focus on stars of solar to super-solar metallicities with 69 models for $Z=$0.014, 0.02, and 0.03, which are expected to have contributed to the molecular cloud where the Sun was born and have been suggested to produce higher variations in the first than in the second $s$-process peak elements. To discuss the broader trend with metallicity, we also present four and two models of $Z=0.04$ and $Z=0.05$, respectively, and a further two models of $Z=0.0028$ and three models of $Z=0.007$, from \cite{karakas18} and \cite{karakas18}, respectively, all re-run with the Set 2 nuclear network of \cite{szanyi25}.

As shown in the tables, depending on mass, metallicity and $N_\mathrm{ov}$, the models can reach C/O$>$1, therefore becoming able to produce dust such as SiC, or remain O-rich (C/O$<$1). Specifically, at $Z=0.007$ and 0.014 all the models are C-rich at the end of the evolution, except for the 2 \msun\ $Z=0.014$ models with $N_\mathrm{ov}$=0. As the metallicity increases, we find a larger fraction of O-rich models because the initial abundance of oxygen is higher and the third dredge-up efficiency usually decreases. Nevertheless, at $Z=0.02$, all the models still become C-rich, except for the 2 \msun\ star with $N_\mathrm{ov}$=0 and 1. At $Z=0.03$ a stronger convective overshoot is required to reach C/O$>1$, as previously found \citep{karakas14Herichmodels}. At this metallicity, all the models with $N_\mathrm{ov}$=0 are O-rich, and increasingly higher values of $N_\mathrm{ov}$ are needed to produce C-rich stars decreasing the initial mass: $N_\mathrm{ov}$=1, 2, and 3 is needed for M=3.5, 3, and 2.5 \msun, respectively. For $Z=0.04$ and $Z=0.05$, none of the models are C-rich, despite their relatively high $N_\mathrm{ov}$ values.

\begin{figure}
\centering
  \includegraphics[width=10 cm]{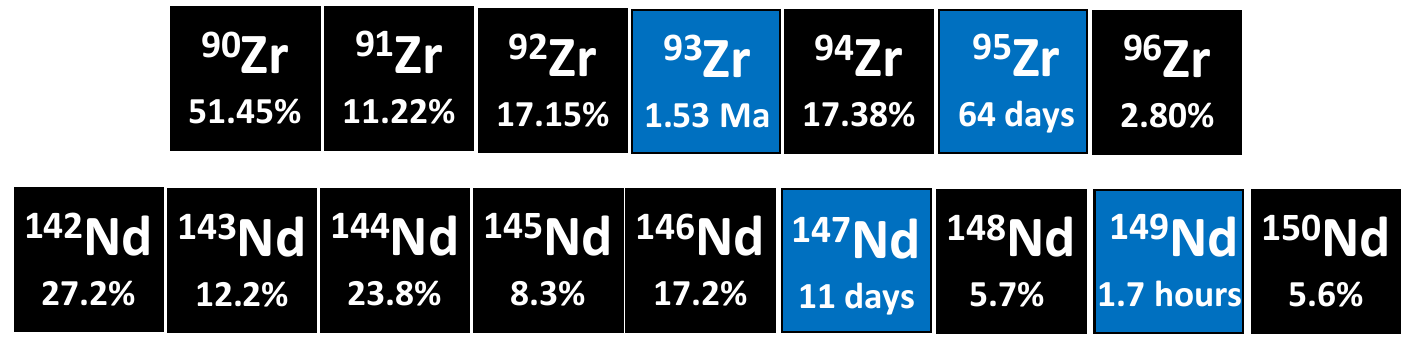}
\caption{Selected sections of the nuclide chart showing the stable isotopes (black boxes, with the contribution of each isotope to the elemental abundance in the Solar System) of Zr (top) and of Nd (bottom) and their unstable isotopes (blue boxes, with their half-lives) potentially located on the $s$-process path of neutron captures (which moves from left to right to higher masses by adding neutrons). In the case of Zr, \iso{93}Zr has a temperature-dependent half-life in any case long enough to make it behave like a stable isotope during the $s$ process, while \iso{96}Zr has a constant half-life and is a branching point with a typically higher probability of decaying than capturing a neutron. In the case of Nd, both \iso{147}Nd and \iso{149}Nd have almost constant half-lives, and live short enough that they typically decay before capturing a neutron. For a more detailed figure, also covering the branching points at the isotopes of Sr and Y that may affect the isotopic composition of Zr, see Figure 1 of \citep{lugaro03grains}.}
\label{fig:chart}       
\end{figure}

\subsection{Calculation of the diluted isotopic compositions and the $\varepsilon$ values}
\label{sec:norm}

We considered the 80 models listed in Tables~\ref{tab:models0p014}, \ref{tab:models0p02}, and \ref{tab:models0p03} and transformed the surface abundances of the Zr and Nd isotopes into the $\varepsilon$ values, described below and needed to compare to the data, using the method described in, e.g., \cite{lugaro23}. In brief: 
\begin{enumerate}
    \item We dilute the AGB surface abundances and add them to the solar abundances. 
This step simulates the mixing of AGB ejecta (in the form of stardust) with material of isotopic composition corresponding to that of the bulk, standard, solar material. We use exactly the same solar distribution used as the initial in the AGB models \citep{asplund09}, because using a different distribution could also results in variations, but not due to the AGB nucleosynthesis effects that we aim to study here \citep{lugaro23}. Note that in the models this distribution is scaled according to the metallicity, while in this Step 1. it is taken to represent the solar matter, therefore it is not scaled. We dilute the abundances by a factor of 10$^4$ (i.e., we multiply them by 0.0001), which corresponds to the order of magnitude of the abundance of stardust SiC grains in meteorities of $\sim$100 parts per million \citep{davidson14,nittler21,nguyen23}. 
\item We calculate the isotopic ratios of interest, \iso{96}Zr/\iso{90}Zr and \iso{148}Nd/\iso{144}Nd, and `internally normalise' them with a second isotopic ratio, \iso{94}Zr/\iso{90}Zr and \iso{146}Nd/\iso{144}Nd, respectively, using the exponential formula with the mass-dependent exponent (Eq.~\ref{eq:2}). This step is needed to factor out the effect of mass-dependent fractionation \citep{lugaro23}, as done for the data to which we compare the models.  
\item We calculate the per ten thousand (i.e., $\times$ 10,000) variation of the final ratio relative to the bulk, standard, solar corresponding ratio, again, the same used as the initial composition of the AGB models.
\end{enumerate}

\noindent For example:

\begin{equation}
    \varepsilon ^{96}\rm{Zr} = {\left[
    {\left(
    \frac{(^{96}\rm{Zr}/^{90}\rm{Zr})_{\it DIL}}
    {(^{96}\rm{Zr}/^{90}\rm{Zr})_{\it{solar}}}
    \right)}
    {\left(
    \frac{(^{94}\rm{Zr}/^{90}\rm{Zr})_{\it{DIL}}}
    {(^{94}\rm{Zr}/^{90}\rm{Zr})_{\it{solar}}}
    \right)}^{\it{-Q}} - 1
    \right]} \times 10^4,
    \label{eq:1}
\end{equation}

\noindent where

\begin{equation}
    Q = \frac{\ln{(96)} - \ln{(90)}}{\ln{(94)} - \ln{(90)}}.
    \label{eq:2}
\end{equation}

\noindent For $\varepsilon$\iso{148}Nd the formula is equivalent, except using \iso{144}Nd as the reference isotope, \iso{146}Nd/\iso{144}Nd as the normalising ratio, and the masses in Eq.~\ref{eq:2} changed accordingly.

In terms of the choice of isotopes, the main isotopes in the numerator are \iso{96}Zr and \iso{148}Nd, i.e., two isotopes not generally produced by the $s$ process, being preceded in the neutron-capture chain by an unstable isotope, \iso{95}Zr and \iso{147}Nd, respectively (Figure~\ref{fig:chart}). All the other isotopes involved in the calculations (\iso{90,94}Zr and \iso{144,146}Nd) are on the main $s$-process path and are typically strongly produced by the $s$ process. This is because they all have an even number of nucleons (\iso{90}Zr has also a magic number of neutrons), which results in relatively small cross sections for neutron captures and therefore high $s$-process production. This choice of isotopes means that a typical $s$-process signature results in negative $\varepsilon$\iso{96}Zr and $\varepsilon$\iso{148}Nd values, whose ratio will necessarily be a positive number, with a value depending on the absolute value of each $\varepsilon$, i.e., on the strength of the $s$-process signature of each element. As a consequence, the value of the $\varepsilon$\textsuperscript{96}Zr/$\varepsilon$\textsuperscript{148}Nd ratio depends on the number of free neutrons at the $s$-process site. The smaller the number of neutrons, the more the first $s$-process peak is reached and Zr will show the strongest $s$-process signature (i.e., $\varepsilon$\textsuperscript{96}Zr will be more negative). The larger the number of neutrons, the more the second $s$-process peak is reached and Nd will show the strongest $s$-process signature (i.e., $\varepsilon$\iso{148}Nd will be more negative). The ratio of the two is therefore a measure of the total number of free neutrons, or more precisely, the total time-integrated neutron flux (so-called neutron exposure). 

This analysis applies only if the neutron density is relatively low, below $\sim 10^9$ cm$^{-3}$. Above this value the branching points at \iso{95}Zr and \iso{147}Nd are activated, \iso{96}Zr and \iso{148}Nd are produced, the $\varepsilon$-values can shift towards positive values, and no longer represent the total number of free neutrons, but rather the neutron density at the $s$-process site. This happens in a few of the AGB models presented here and will be discussed below for these specific cases. In any case, AGB stardust SiC data show depletions in \iso{96}Zr and \iso{148}Nd, relative to \iso{94}Zr and \iso{144}Nd, respectively \citep[e.g.][]{richter92,liu14Zr}. Therefore, the possible signature in meteoritic bulk rocks of the SiC stardust from AGB stars already recovered in the meteoritic stardust inventory will correspond to AGB models where \iso{96}Zr and \iso{148}Nd are depleted, rather than produced. 

The observed $\varepsilon$\textsuperscript{96}Zr/$\varepsilon$\textsuperscript{148}Nd ratio can be determined as the slope of the data from bulk chondrites by plotting $\varepsilon$\textsuperscript{96}Zr against $\varepsilon$\textsuperscript{148}Nd. Specifically, the available data for Nd \citep{burkhardt16,fukaiyokoyama17,yokoyamafukai19,saji20,frossard22} and Zr \citep{akram15bulkZr,render22} were compiled and weighted averages for different chondrite groups (OC, EC, CO, CM, CV, CR) were calculated for each isotope. In general, the $\varepsilon$\textsuperscript{96}Zr is $\sim$10 times larger than $\varepsilon$\textsuperscript{148}Nd, therefore, we used a $\varepsilon$\textsuperscript{96}Zr/$\varepsilon$\textsuperscript{148}Nd ratio of $\sim$10 for comparison with the stellar models. Note that this slope is currently not well defined because published data for Zr and Nd in carbonaceous chondrites are sparse. Nevertheless, new and unpublished data for Nd \citep{ball24,ball25} and Zr (Ball et al., in preparation) measured on the same sample aliquots support this value.

\begin{table}
\caption{The 2, 3, and 26 AGB models of $Z=0.0028$, $Z=0.007$, and $Z=0.014$, respectively, are listed including their mass ($M$), mass of the partial mixing zone ($M_{\rm mixed}$), overshoot parameter ($N_\mathrm{ov}$), and final C/O ratios. We also report the $\varepsilon$\iso{96}Zr and $\varepsilon$\iso{148}Nd values calculated from the final surface abundances using a dilution factor of 10$^4$ and their ratios (also plotted in Figures~\ref{fig:ratios} and \ref{fig:choice}). 
The SL label indicates the models that reach the C/O$>$1 condition for the lowest value of $N_\mathrm{ov}$ for each mass and metallicity, which we selected to be included in the summarising Figure~\ref{fig:choice}. Unusual values discussed further in the text are highlighted by asterisks.}
\label{tab:models0p014}
\begin{tabular}{lllcccc}
\hline\noalign{\smallskip}
 $M$ (\msun) & $M_{\rm mixed}$ (\msun) & $N_\mathrm{ov}$ & C/O & $\varepsilon$\iso{96}Zr  & $\varepsilon$\iso{148}Nd & $\varepsilon$\iso{96}Zr/$\varepsilon$\iso{148}  \\
 \noalign{\smallskip}\hline\noalign{\smallskip}
  \multicolumn{7}{c}{$Z=0.00028$} \\
  \hline\noalign{\smallskip}
3	&	$2 \times 10^{-3}$	&	0	&	9.8	&	35	&	$-$19	&	$-$1.8	\\
4	&	$1 \times 10^{-3}$	&	0	&	6.3	&	43	&	$-$12	&	$-$3.6	\\
\noalign{\smallskip}\hline\noalign{\smallskip}
  \multicolumn{7}{c}{$Z=0.007$} \\
  \hline\noalign{\smallskip}
1.75	&	$2 \times 10^{-3}$	&	0	&	2.0	&	$-$10	&	$-$27	&	0.4	\\
3   &	$2 \times 10^{-3}$	&	0	&	5.1	&	$-$26	&	$-$27	&	1.0	\\
4	&	$1 \times 10^{-3}$	&	0	&	3.3	&	99	&	$-$18	&	$-$5.3	\\
\hline\noalign{\smallskip}
  \multicolumn{7}{c}{$Z=0.014$} \\
  \hline\noalign{\smallskip}
2	&	$1 \times 10^{-3}$	&	0	&	0.6	&	$-$4.1	&	$-$3.9	&	1.1	\\
	&		&	1 SL 	&	1.1	&	$-$12	&  $-$13	&	1.0	\\
	&	$2 \times 10^{-3}$	&	0	&	0.6	&	$-$7.4	&	$-$6.4	&	1.2	\\
	&		&	1 SL 	&	1.1	&	$-$24	&	$-$21	&	1.1	\\
2.5	&	$1 \times 10^{-3}$	&	0 SL 	&	1.6	&	$-$23	&	$-$15	&	1.5	\\
	&		&	1	&	2.0	&	$-$26	&	$-$18	&	1.5	\\
	&		&	2	&	2.5	&	$-$25	&	$-$17	&	1.5	\\
	&		&	3	&	3.5	&	$-$30	&	$-$20	&	1.5	\\
	&	$2 \times 10^{-3}$	&	0 SL 	&	1.5	&	$-$39	&	$-$25	&	1.6	\\
	&		&	1	&	1.9	&	$-$46	&	$-$30	&	1.5	\\
	&		&	2	&	2.3	&	2.7*	&	$-$76	&	$-$0.03*	\\
	&		&	3	&	3.2	&	$-$11	&	$-$73	&	0.2* \\
3    &	$5 \times 10^{-4}$	&	0 SL 	&	2.5	&	$-$16	&	$-$11	&	1.4	\\
	&	$1 \times 10^{-3}$	&	0 SL 	&	2.4	&	$-$31	&	$-$16	&	2.0	\\
	&		&	1	&	2.8	&	$-$36	&	$-$19	&	1.9	\\
	&		&	2	&	3.4	&	$-$39	&	$-$19	&	2.1	\\
	&		&	3	&	4.0	&	$-$42	&	$-$16	&	2.6	\\
	&	$2 \times 10^{-3}$	&	0 SL 	&	2.2	&	$-$51	&	$-$25	&	2.0	\\
	&		&	1	&	2.7	&	$-$56	&	$-$28	&	2.0	\\
	&		&	2	&	3.2	&	$-$62	&	$-$29	&	2.1	\\
	&		&	3	&	3.8	&	$-$66	&	$-$28	&	2.4	\\
	&		&	4	&	6.1	&	$-$83	&	$-$25	&	3.3	\\
3.5	&	$1 \times 10^{-3}$	&	0 SL 	&	2.5	&	$-$25	&	$-$9.8	&	2.5	\\
	&		&	1	&	2.5	&	$-$26	&	$-$10	&	2.6	\\
	&		&	2	&	2.6	&	$-$26	&	$-$9.5	&	2.7	\\
	&	$2 \times 10^{-3}$	&	0 SL 	&	2.3	&	$-$31	&	$-$17	&	1.8	\\
\noalign{\smallskip}\hline\noalign{\smallskip}
\end{tabular}
\end{table}

\begin{table}
\caption{Same as Table~\ref{tab:models0p014} but for the 23 AGB models of $Z=0.02$.}
\label{tab:models0p02}
\begin{tabular}{lllcccc}
\hline\noalign{\smallskip}
 $M$ (\msun) & $M_{\rm mixed}$ (\msun) & $N_\mathrm{ov}$ & C/O & $\varepsilon$\iso{96}Zr  & $\varepsilon$\iso{148}Nd & $\varepsilon$\iso{96}Zr/$\varepsilon$\iso{148}  \\
\noalign{\smallskip}\hline\noalign{\smallskip}
2	&	$1 \times 10^{-3}$ &	1	&	0.5	&	$-$3.3	&	$-$1.1	&	3.1	\\
	&		&	2 SL 	&	1.1	&	$-$20	&	$-$7.2	&	2.7	\\
	&		$2 \times 10^{-3}$ &	1	&	0.5	&	$-$6.3	&	$-$2.9	&	2.2	\\
	&		&	2 SL 	&	1.1	&	$-$34	&	$-$36	&	0.9	\\
2.5	&	$1 \times 10^{-3}$	&	0 SL 	&	1.1	&	$-$25	&	$-$8.3	&	3.0	\\
	&		&	1	&	1.4	&	$-$29	&	$-$10	&	2.9	\\
	&	$2 \times 10^{-3}$	&	0 SL 	&	1.1	&	$-$40	&	$-$17	&	2.3	\\
	&		&	1	&	1.4	&	$-$51	&	$-$22	&	2.4	\\
3   &	$5 \times 10^{-4}$	&	0 SL 	&	1.7	&	$-$22	&	$-$4.7	&	4.6	\\
	&	$1 \times 10^{-3}$	&	0 SL 	&	1.7	&	$-$36	&	$-$10	&	3.7	\\
	&		&	1	&	1.9	&	$-$38	&	$-$11	&	3.5	\\
	&		&	2	&	2.4	&	$-$45	&	$-$11	&	4.2	\\
	&		&	3	&	3.3	&	$-$50	&	$-$9.4	&	5.4	\\
	&		&	4	&	6.0	&	$-$70	&	$-$10	&	7.0	\\
    &	$2 \times 10^{-3}$	&	0 SL 	&	1.6	&	$-$50	&	$-$18	&	2.7	\\
	&		&	1	&	1.8	&	$-$54	&	$-$21	&	2.6	\\
	&		&	2	&	2.3	&	$-$64	&	$-$23	&	2.8	\\
	&		&	3	&	3.2	&	$-$76	&	$-$22	&	3.5	\\
	&		&	4	&	5.7	&	$-$108	&	$-$25	&	4.3	\\
3.5	&	$1 \times 10^{-3}$	&	0 SL 	&	1.7	&	$-$30	&	$-$7.3	&	4.0	\\
	&		&	1	&	1.8	&	$-$32	&	$-$7.5	&	4.2	\\
	&		&	3	&	1.3	&	$-$21	&	$-$4.8	&	4.3	\\
	&	$2 \times 10^{-3}$	&	0 SL 	&	1.6	&	$-$41	&	$-$15	&	2.8	\\
\noalign{\smallskip}\hline\noalign{\smallskip}
\end{tabular}
\end{table}

\begin{table}
\caption{Same as Table~\ref{tab:models0p014} but for the 20, 4, and 2 AGB models of $Z=0.03$, 0.04, and 0.05, respectively. 
}
\label{tab:models0p03}
\begin{tabular}{lllcccc}
\hline\noalign{\smallskip}
 $M$ (\msun) & $M_{\rm mixed}$ (\msun) & $N_\mathrm{ov}$ & C/O & $\varepsilon$\iso{96}Zr  & $\varepsilon$\iso{148}Nd & $\varepsilon$\iso{96}Zr/$\varepsilon$\iso{148}Nd  \\
\noalign{\smallskip}\hline\noalign{\smallskip}
  \multicolumn{7}{c}{$Z=0.03$} \\
  \hline\noalign{\smallskip}
2.5	&	$1 \times 10^{-3}$	&	0	&	0.4	&	$-$0.4	&	$-$0.1	&	5.6 \\
	&		&	1	&	0.5	&	$-$8.9	&	$-$2.1	&	4.2	\\
	&		&	2	&	0.8	&	$-$22	&	$-$4.5	&	4.9	\\
	&		&	3 SL 	&	1.5	&	$-$40	&	$-$7.7	&	5.2	\\
	&	$2 \times 10^{-3}$	&	0	&	0.4	&	$-$0.7	&	$-$0.2	&	3.6 \\
	&		&	1	&	0.5	&	$-$15	&	$-$5.8	&	2.6	\\
	&		&	2	&	0.8	&	$-$22	&	$-$4.5	&	4.9	\\
3    &	$5 \times 10^{-4}$	&	2 SL 	&	1.2	&	$-$19	&	$-$2.4	&	8.0	\\
     &                      & 4 & 2.0 & $-$17 & $-$1.3 & 13 \\
    &	$1 \times 10^{-3}$	&	0	&	0.6	&	$-$15	&	$-$4.0	&	3.8	\\
	&		&	1	&	0.8	&	$-$23	&	$-$5.5	&	4.3	\\
	&		&	2 SL 	&	1.1	&	$-$31	&	$-$6.5	&	4.9	\\
	&		&	4	&	1.9	&	$-$31	&	$-$4.3	&	7.3	\\
	&	$2 \times 10^{-3}$	&	0	&	0.6	&	$-$19	&	$-$7.9	&	2.4	\\
	&		&	1	&	0.8	&	$-$29	&	$-$11	&	2.7	\\
	&		&	2 SL 	&	1.1	&	$-$40	&	$-$13	&	3.0	\\
	&		&	4	&	1.8	&	$-$47	&	$-$10	&	4.6	\\
3.5	&	$1 \times 10^{-3}$	&	0	&	0.9	&	$-$20	&	$-$4.5	&	4.4	\\
	&		&	1 SL 	&	1.0	&	$-$24	&	$-$4.9	&	5.0	\\
	&	$2 \times 10^{-3}$	&	1 SL 	&	1.0	&	$-$34	&	$-$9.7	&	3.5	\\
    \noalign{\smallskip}\hline\noalign{\smallskip}
  \multicolumn{7}{c}{$Z=0.04$} \\
  \hline\noalign{\smallskip}
3    &	$1 \times 10^{-3}$	&	2  	&	0.50 &	$-$9.9		&	$-$2.2	&	4.5	\\ 
    &		&	3  	&	0.76 &	$-$20		&	$-$3.6	&	5.5	\\  
3    &	$2 \times 10^{-3}$	&	2  	&	0.49 &	$-$14		&	$-$4.7	&	3.0	\\
    &		&	3  	&	0.73 &	$-$29		&	$-$9.1	&	3.2	\\ 
  \noalign{\smallskip}\hline\noalign{\smallskip}
  \multicolumn{7}{c}{$Z=0.05$} \\
  \hline\noalign{\smallskip}
3    &	$1 \times 10^{-3}$	&	3 	&	0.40 &	$-$2.5	&	$-$0.4	&	6.4	\\
    &		&	4  	&	0.62 &	$-$14		&	$-$2.0	&	6.8	\\ 
\noalign{\smallskip}\hline\noalign{\smallskip}
\end{tabular}
\end{table}

\section{Results and discussion}
\label{sec:results}

\begin{figure}
\centering
  \includegraphics[width=10 cm]{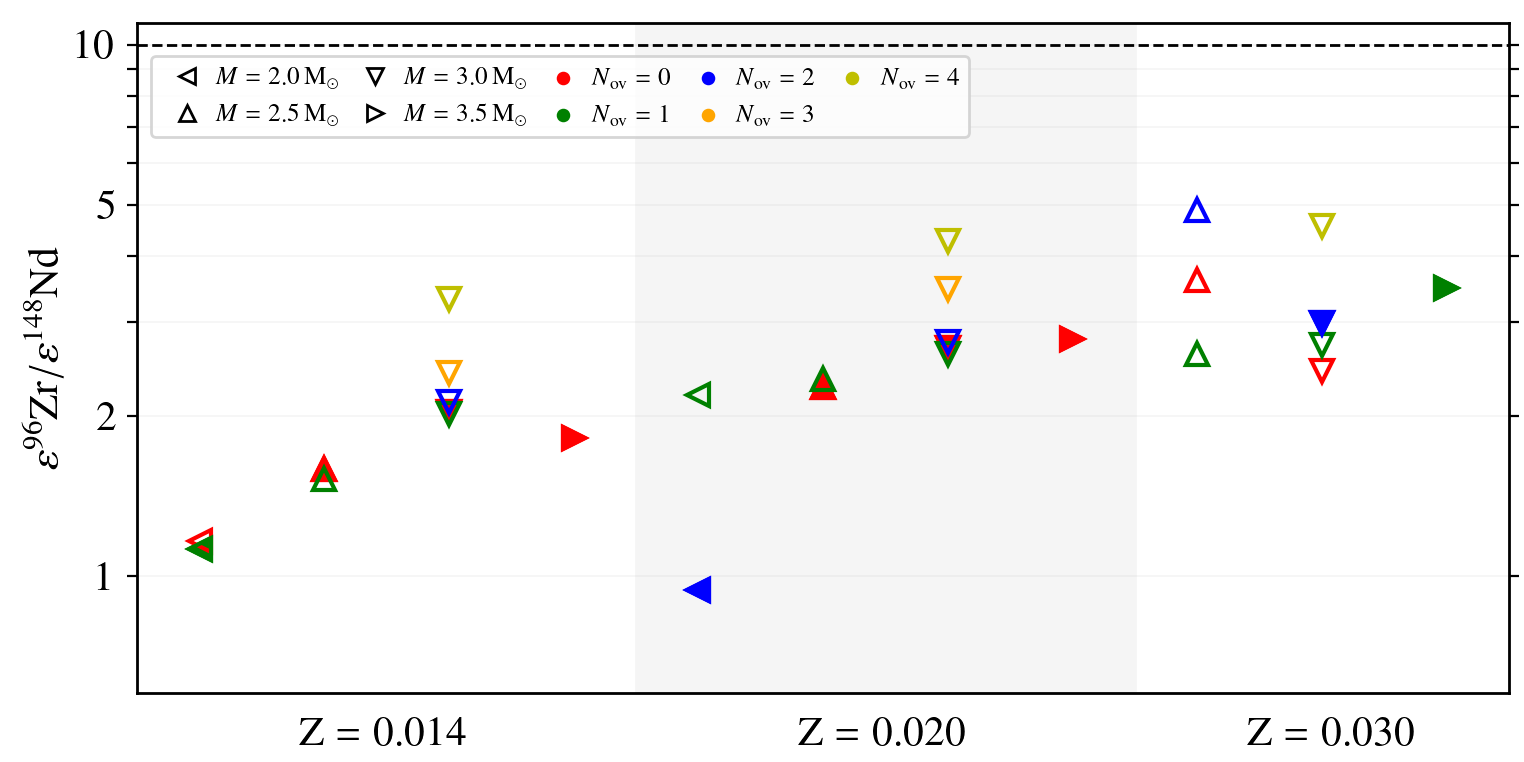}
  \includegraphics[width=10 cm]{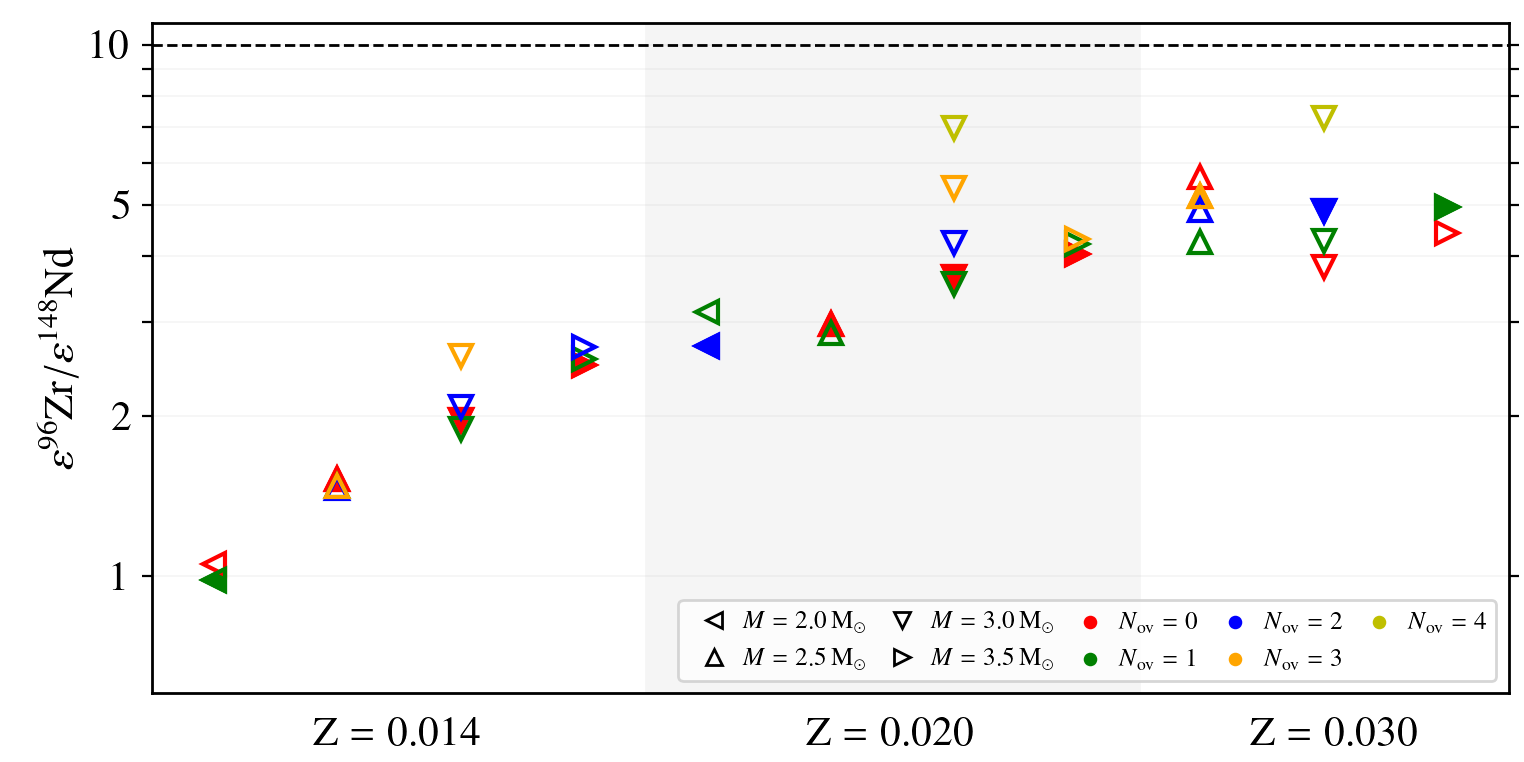}
\caption{
The predicted $\varepsilon$\iso{96}Zr to $\varepsilon$\iso{148}Nd ratios calculated from the final surface abundances of the AGB star models of metallicities $Z$ indicated on the x-axis, initial stellar masses between 2 and 3.5 \msun\ ($M$, different symbols, as indicated in the legend box), variable values of the $N_\mathrm{ov}$ parameter (different colors, as indicated in the legend box) and fixed mass of the mixed region $M_{\rm mixed} = 2 \times 10^{-3}$ \msun\ (top) and $1 \times 10^{-3}$ \msun\ (bottom). The full symbols correspond to the SL models plotted in Figure~\ref{fig:choice}. The observed ratio of the order of $\sim$10 is indicated as a dashed horizontal line.}
\label{fig:ratios}       
\end{figure}
\begin{figure}
\centering
  \includegraphics[width=10 cm]{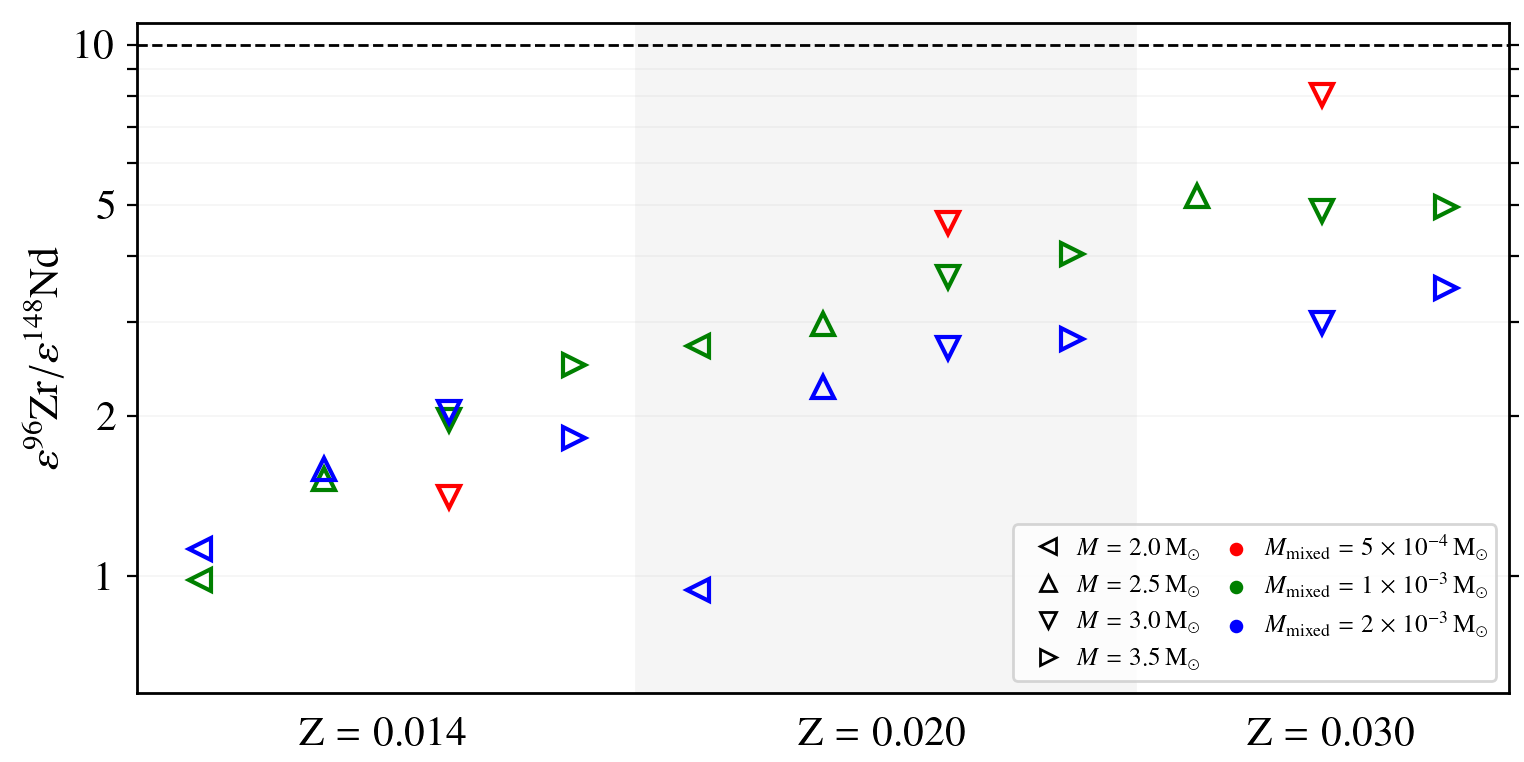}
\caption{Same as Figure~\ref{fig:ratios}, but for the selected SL models only, i.e., with one value of $N_\mathrm{ov}$ for each given mass, metallicity, and $M_{\rm mixed}$.}
\label{fig:choice}       
\end{figure}

Tables \ref{tab:models0p014}, \ref{tab:models0p02}, and \ref{tab:models0p03} list the $\varepsilon$\iso{96}Zr, $\varepsilon$\iso{148}Nd, and their ratio calculated for each model. Figures~\ref{fig:ratios} and \ref{fig:choice} show the $\varepsilon$\iso{96}Zr/$\varepsilon$\iso{148}Nd ratio from all the $Z=0.014$, 0.02, and 0.03 models with $M_{\rm mixed} = 1 \times 10^{-3}$ \msun\ and $2 \times 10^{-3}$ \msun, and the selected SL models with $M_{\rm mixed} = 5 \times 10^{-3}$ \msun\ 
- except for the two 2.5 \msun, $Z=0.014$, $M_{\rm mixed}=2 \times 10^{-3}$ \msun\ models with ratios 0 and $0.2$, which will be discussed below. All the values reported in the tables and figures mentioned above are the final values at the end of the stellar evolution, noting that most of the mass loss occurs at the end of the evolution, so we expect these values to be the predominant signature in the forming dust. The absolute $\varepsilon$\iso{96}Zr and $\varepsilon$\iso{148}Nd values reported in the tables depend on the absolute abundance of each isotope and linearly on the chosen dilution factor (for example, a 10 times higher/lower dilution produces $\varepsilon$ values ten times lower/higher), which we keep constant at $10^4$. 
Their ratio, instead, is independent of the dilution factor
(it remains constant within 7\% considering all our models when changing the dilution by a factor of 10 upward and downward), 
as far as there is a enough material to produce an $s$-process signature. 
Otherwise, the value of the ratio does not represents a physically carried signature, but just the numerical results of the division of very small numbers.




We consider three values for the size of the mass of partial mixing zone ($M_{\rm mixed}$) leading to the formation of the \iso{13}C nuclei that represent the main neutron source via the \iso{13}C($\alpha$,n)\iso{16}O reaction. Note that, as described in detail in, e.g., \cite{buntain17}, only the bottom roughly half of $M_{\rm mixed}$, i.e., the region with the lower number of protons relative to \iso{12}C results in the \iso{13}C-rich pocket, efficiently producing the $s$-process elements. In the top part of $M_{\rm mixed}$, instead, the higher number of protons relative to \iso{12}C results in proton captures also on \iso{13}C itself, and the formation of \iso{14}N instead.
As in our computational tools $M_{\rm mixed}$ is included in a parametrised way during the post-processing, we cannot study the possible dependence between its value and $N_\mathrm{ov}$, therefore, we tested our typical range of $M_{\rm mixed}$.
According to \cite{karakas16}, for the stellar mass range considered here, $M_{\rm mixed}$ of the order of 1-2 $\times 10^{-3}$ \msun\ is required to match the observational constraint that typical galactic AGB stars (of metallicity around solar or lower) are enhanced in $s$-process elements by up to a factor of $\sim$10. All the models with $M_{\rm mixed}$ in this range show $s$-process production efficient enough to result in $\varepsilon$ variations above unity when using a dilution of 10$^4$, except for the 2.5 \msun, $Z=0.03$ model with $N_\mathrm{ov}$=0.

Figure~\ref{fig:ratios} shows that the $\varepsilon$\iso{96}Zr/$\varepsilon$\iso{148}Nd ratios are within a factor of 2 for any given mass, metallicity, and $M_{\rm mixed}$ when varying $N_\mathrm{ov}$ and they increase with increasing metallicity and mass. The positive trend with metallicity is expected because in higher-metallicity stars it is more difficult for the $s$ process to reach the second peak as the number of neutrons captured by each Fe seed is lower. This is because these stars contain more Fe seeds than their lower-metallicity couterparts but the same amount of free neutrons, given that \iso{13}C does not depend on the metallicity of the star (as explained in Section~\ref{sec:data}). The number of neutrons captured by each Fe seed is lower at higher metallicity, therefore, it is more difficult for the $s$-process path to reach the second peak. 
This trend is confirmed when extending the metallicity: the $Z=0.007$ models have generally lower $\varepsilon$\iso{96}Zr/$\varepsilon$\iso{148}Nd ratios than the $z=0.014$ models, except for the 4 \msun\ model. This model shows a significant activation of the \iso{22}Ne($\alpha$,n)\iso{22}Ne neutron source, due to the maximum temperature in the thermal pulses reaching 360 MK, which results in positive $\varepsilon$\iso{96}Zr, as expected by the activation of the \iso{95}Zr branching point. This results in a negative slope, as opposed to the observations. Decreasing the metallicity down to $Z=0.0028$ highlights this effect even more, as the stars become even hotter.
Increasing the metallicity does not result in a significant variations of the results, in any case, the $z=0.04$ and the $z=0.05$ models do not become C-rich, so they could not have produced SiC grains. Moreover, they are not plausible as source of the observed anomalies as the maximum Fe enrichment observed in stars of ages similar to the Sun is around a factor of two \cite{casagrande11,nissen20}.

\begin{figure}
\centering
  \includegraphics[width=10 cm]{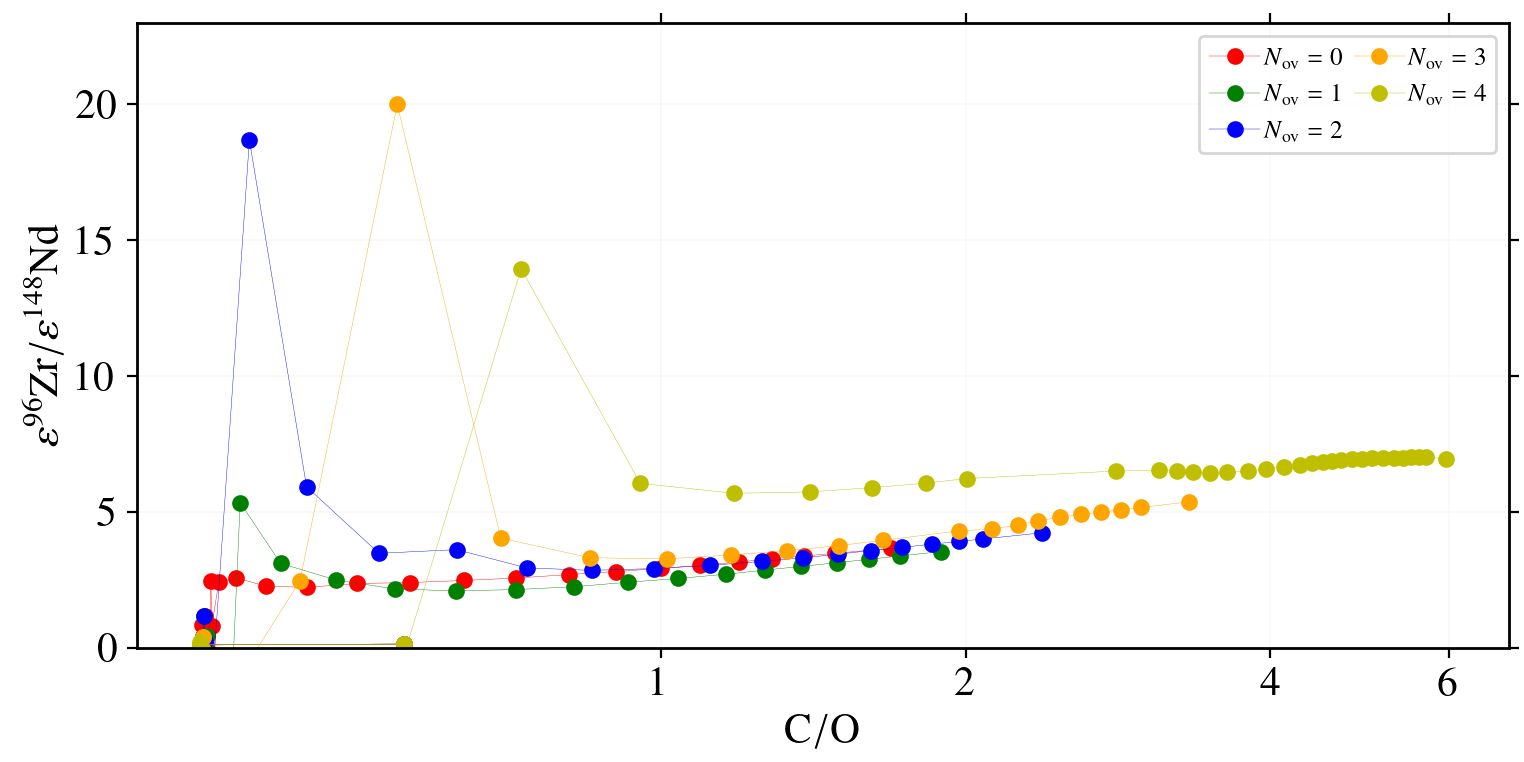}
    \includegraphics[width=10 cm]{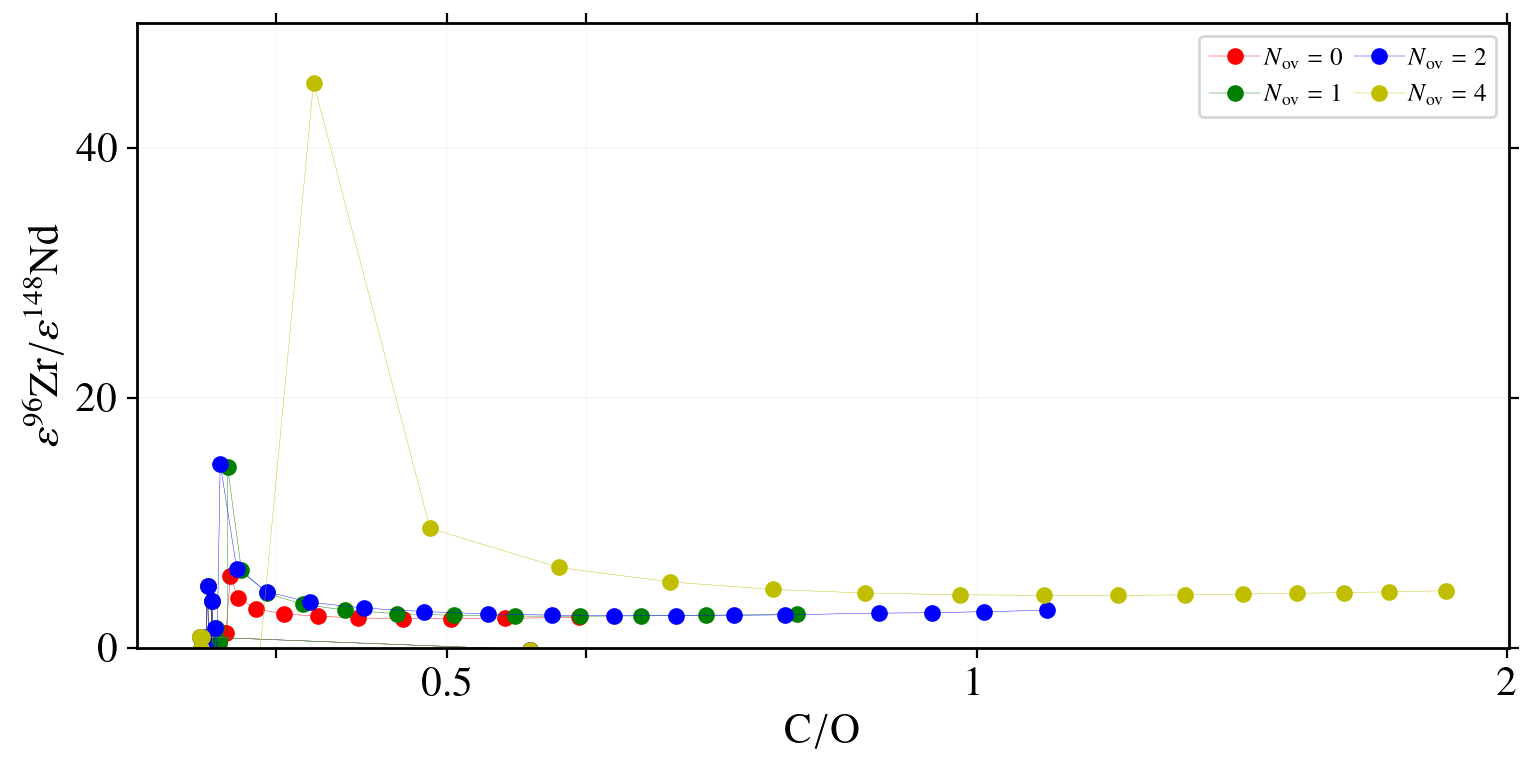}
\caption{The $\varepsilon$\iso{96}Zr/$\varepsilon$\iso{148}Nd ratio as function of the C/O ratio calculated using the Zr and Nd isotopic abundances as they evolve at the surface of the 3 \msun, $Z=0.02$ models with $M_{\rm mixed}= 1 \times 10^{-3}$ \msun\ (top) and the 3 \msun, $Z=0.03$ models with $M_{\rm mixed}= 2 \times 10^{-3}$ \msun\ (bottom).}
\label{fig:evolution}       
\end{figure}

Apart from the $Z=0.0028$ and the 4 \msun, $Z=0.007$ models, there is no indication of a strong effect of the \iso{22}Ne($\alpha$,n)\iso{25}Mg reaction in any other models. We tested this by running the 9 selected (SL) 3 \msun\ models with the rate of this reaction set to zero, and we did not find differences in the resulting $\varepsilon$\iso{96}Zr/$\varepsilon$\iso{148}Nd beyond 30\%. Therefore, we conclude that the trend with mass is due to the neutron exposure decreasing in the \iso{13}C pocket as the star evolves, and the abundances of the nuclei heavier than iron increase, thereby capturing more free neutrons, as shown already by Gallino et al. (1998) \cite[][see their Figure 6]{gallino98}. This behaviour is also noticeable in the top panel of Figure~\ref{fig:evolution}, which we will discuss later. The effect is stronger for higher masses given their longer lifetimes and higher number of \iso{13}C pockets, e.g., 7 versus 20, for the 2 \msun\ and the 3.5 \msun\ models at $Z=0.014$. However, it becomes weaker for higher $Z$ both because the difference in the number of \iso{13}C pockets is smaller, e.g., 12 versus 19, for the 2.5 \msun\ and the 3.5 \msun\ models at $Z=0.03$, and because these models already start with higher abundances, so the impact of increasing the abundances of the nuclei heavier than iron is relatively smaller.


The trend with $N_\mathrm{ov}$ is more complex as it is related to the impact of the neutrons released by the \iso{13}C nuclei early during the AGB evolution when the star is relatively cool and not all the \iso{13}C nuclei burn during the radiative interpulse period. The left-over \iso{13}C nuclei are ingested in the following thermal pulse \citep[see also discussion in][]{lugaro14Zr}, where they produce a neutron flux with higher densities and fewer free neutrons than in the case when the \iso{13}C nuclei burn radiatively. This is because the \iso{13}C burns faster, as the temperatures are higher, but together with \iso{14}N, which is also ingested both from the H-burning ashes and the top region of the partial mixing zone, and it is a neutron poison via the \iso{14}N(n,p)\iso{14}C reaction. These ingestion events favour the production of the first peak relative to the second peak, resulting in higher $\varepsilon$\iso{96}Zr/$\varepsilon$\iso{148}Nd ratios. Some example results are shown in Figure~\ref{fig:evolution}, where it is clear that the $N_\mathrm{ov}$ parameter strongly controls this effect. This is expected, as higher $N_\mathrm{ov}$ results in the third dredge-up occurring at lower core masses, therefore, the partial mixing zone and the \iso{13}C pocket will also occur at lower core masses, when the star is cooler, thus favouring \iso{13}C ingestions. The models with the highest $N_\mathrm{ov}$ retain the memory of these events until the end of the evolution and therefore have higher ratios in general. This effect is more prominent at higher metallicities, as the star is generally cooler, and for smaller $M_{\rm mixed}$ because the \iso{13}C-rich region is located closer to the top of the He-rich region, therefore, at lower temperatures. 

The two models corresponding to 2.5 \msun\ stars at $Z=0.014$, $M_{\rm mixed}= 2 \times 10^{-3}$ \msun\ and $N_\mathrm{ov}$=2 and 3, with close to zero and zero ratios, respectively (indicated by asterisks in Table~\ref{tab:models0p014})
are special because they produce excesses in \iso{96}Zr relative to, e.g., \iso{94}Zr and to solar, rather than the depletions typical of the $s$ process. 
These excesses are again produced early during the AGB evolution when the \iso{13}C nuclei are ingested in the following thermal pulse. In these models, the neutron density is high enough in these phases to activate the \iso{95}Zr branching point, unlike 
the typical $s$-process Zr signature. In the most extreme case, we even obtain a positive $\varepsilon$\iso{96}Zr. While this type of AGB signature has not been observed in any stardust SiC grains from AGB stars so far, we show it here as it is possible that some grains with \iso{96}Zr excesses may have been misclassified as supernova grains, and/or that other types of AGB grains may have carried this signature in the early Solar System\footnote{We did not include 2 \msun, $Z=0.014$ models with $N_\mathrm{ov}$=2 in this paper, which also present such an effect, even much more pronounced with excesses in \iso{96}Zr of a factor of $\sim$3 relative to \iso{94}Zr and to solar and $\varepsilon$\iso{96}Zr reaching almost +1000. We do not discuss these models here because the strong \iso{13}C ingestion may affect the energy generation and the stellar structure \cite{bazan93}. More investigation is needed, which we 
cannot do with our current stellar structure code because it does not include the \iso{13}C pocket (as this is added later in the post-processing), nor the \iso{13}C($\alpha$,n)\iso{16}C reaction, because it does not normally produce significant energy.}.

Overall, higher metallicities result in higher ratios, and the maximum values are reached with the highest values of $N_\mathrm{ov}$ and the lowest values of $M_{\rm mixed}$. The maximum value of 13 is reached by the 3 \msun\ $Z=0.03$ model with $N_\mathrm{ov}$=4 and $M_{\rm mixed}$ = 5 $\times 10^{-4}$ \msun, whose predictions are generally still compatible with presolar grain data and Ba stars observations.

In terms of absolute values, these best fit models produce $\varepsilon$\iso{96}Zr $>$10, which are much higher than the typical variations $<$10 observed in the data. In any case, values $<$10 can be achieved by increasing the dilution factor so that 10$^5$ of AGB material, instead of 10$^4$, is mixed with the solar abundances at Step 1. described in Section~\ref{sec:norm}. This may indicate that a 10 times smaller fraction of AGB stardust grains present in the meteoritic inventory contributed this signature, or, more likely, that the mass balance of all the AGB stardust that carried Zr and Nd into the Solar System resulted in this signal. 
This mass balance should include, for example, the abundant sub-micron size grains ($\sim$0.2 $\mu$m on average) expected to originate from AGB stars of $\sim$ solar metallicity \citep{cristallo20}, therefore, carrying a ratio much lower than $\sim 10$. Moreover, $s$-process dust carriers other than SiC, such as oxides and silicates are expected and predicted to have been present on the basis of leachates data 
\citep[e.g.,][]{schonbachler05leachatesZr,reisberg09leachatesOs}, but currently not well constrained. Therefore, the information needed to calculate this mass balance, i.e., the initial grain abundances and size distributions of all types of AGB $s$-process grains and how much of each element was carried by the different grains, is currently uncertain. Moreover, the final results would also depend on the grain distribution in the disk, as affected by processes that could have selectively destroyed and removed different types of grains. 
In summary, resolving the full inventory of stardust carriers from AGB stars present in the early Solar System is a challenging task and, as discussed above, there are still significant uncertainties. Here we have shown that material from high-metallicity AGB stars could explain the observed Zr/Nd ratio in meteorites without chemical fractionation. If and how this material contributed to the isotopic variations seen in meteorites is a topic that requires further study.


These values of metallicity and $M_{\rm mixed}$ also allow us to best match the isotopic composition of Sr, Zr, and Ba in the large ($\mu$m-sized) SiC grains from AGB stars \citep{lugaro18grains,lugaro20,szanyi25}. To further determine the feasibility of these best-fit AGB models, we can also use independent constraints from Ba stars. These stars accreted $s$-process abundances from a former AGB companion and therefore show excesses relative to solar in heavy elements from the first (e.g., Y and Zr) and second (e.g, La and Ce) $s$-process peaks \cite[e.g.,][]{cseh18}. 

\begin{figure}
\centering
  \includegraphics[width=10 cm]{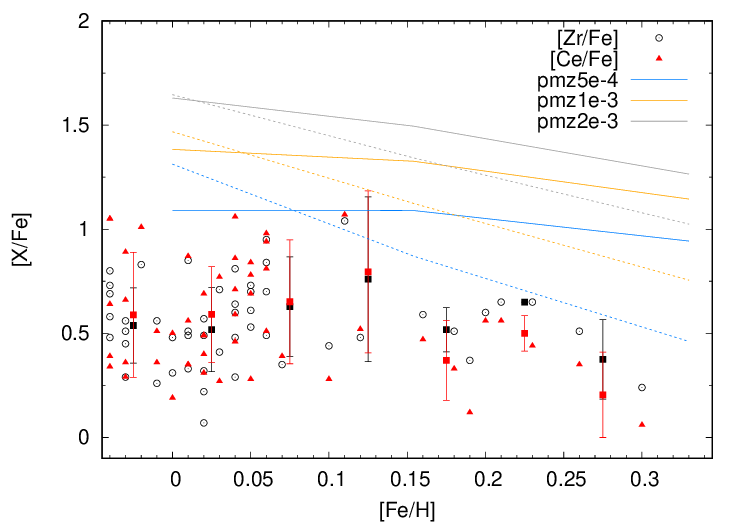}
\caption{Abundances of Zr (black circles) and Ce (red triangles) in giant Ba stars \cite{decastro16,pereira11} with metallicities (represented by the Fe abundance) comparable to the stellar models considered here. The square bracket notation indicates ratios relative to solar, in Log$_{10}$ scale (so that zero is solar). The black and red squares represent the average of the [Zr/Fe] and [Ce/Fe] values, respectively, in 0.05 dex bins in [Fe/H] (with the right edge points included in the averaging). The error bars are the 1$\sigma$ standard deviations, note that the black square at [Fe/H]=0.225 has 0 error, since the two Zr dots have the same value. Overplotted are the [Zr/Fe] and [Ce/Fe] (solid and dashed lines, respectively) predicted at the three metallicities ([Fe/H]=0, 0.15, and 0.33) by the SL 3 \msun\ models plotted in Figure~\ref{fig:choice} for the three different choices of $M_{\rm mixed}$= 0.5, 1, and 2 $\times 10^{-3}$ \msun.}
\label{fig:Bastars}       
\end{figure}

It has been shown and discussed previously \cite[e.g.,][]{cseh18,lugaro20} that the Ce/Zr ratio in Ba stars decreases with increasing stellar metallicity, in agreement with the AGB model predictions given the metallicity effect of the Fe seeds. To expand this previous analysis, in Figure~\ref{fig:Bastars} we compare the observed and predicted Zr and Ce abundances relative to Fe, and to solar. Note that it is not possible to derive the metallicity trend of the $s$ process from this figure because the plotted ratios are affected by the dilution due to mass transfer from the companion, which needs to be considered star by star, and which means that the original $s$-process trend can be lost. 
The trend with metallicity can only be studied using ratios of elements heavier than iron relative to each other, e.g. [Ce/Zr] or [Ce/Y] as done in Figure 1 of \cite{lugaro20}, which includes 180 Ba stars, because the effect of dilution applies to both elements and therefore cancels out (at least in first approximation).  
Moreover, note that in Figure~\ref{fig:Bastars} we focus on the high-metallicity range of Ba star, where the statistics is very low: there are only 8 stars with Fe corresponding to Z=0.02 to 0.03. Therefore, when comparing models and data in Figure~\ref{fig:Bastars}, the only requirement is that the AGB models should produce abundances higher than observed, so that there is some room to decrease these abundances via mass trasfer, potentially by a different dilution factor for each star.

In general, this constrain is satisfied by the AGB models considered here. For example, the smallest $M_{\rm mixed}$ still results in [Zr/Fe]$\sim$1 at $Z=0.02$ and 0.03, 2.5 times above the data, thereby allowing for a dilution of 40\% AGB material to 60\% original Ba-star material. The [Ce/Fe] values instead are closer to the data, and favour values of $M_{\rm mixed}$ above $5 \time 10{-4}$ \msun. The predictions from the models with $N_\mathrm{ov}$=4 do not change the conclusions, as they are up to 0.3 dex higher for Zr, but close to the values plotted for Ce.


\section{Conclusions and future work}
\label{sec:conclusions}



We have compared the $s$-process variability of the first versus second $s$-process peak elements (specifically Zr versus Nd) observed in bulk meteorites with 80 new models of AGB stars of mass between 2 and 3.5 \msun\ and metallicity between solar and twice solar, with variable amounts of the overshoot that controls the efficiency of the third dredge-up, and of the mass of the partial mixing zone (between $5 \times 10^{-4}$ and $2 \times 10^{-3}$ \msun) that leads to the formation of the region rich in the neutron source \iso{13}C. The roughly one order of magnitude difference in the $s$-process variability seen between Zr and Nd can be matched by models of AGB stars of higher-than-solar metallicity and higher $N_\mathrm{ov}$ parameter and/or smaller mass of the partial mixing zone. 
Models of metallicity higher than solar also best match the $s$-process composition of large ($\mu$m-sized) stardust SiC grains from AGB stars recovered from meteorites and their elemental predictions are also in agreement with observations of Ba stars of solar and super-solar metallicity. We conclude that $s$-process nucleosynthesis in AGB stars of metallicity higher than solar can reproduce the observed variations, and no separate dust carriers and associated chemical effects need to be invoked to match the data. 

A population of old stars ($\sim$5-9 Ga) with metallicity higher than solar is present in the galactic solar neighbourhood \cite{nissen20}. If this population was already present at least 4.6 Ga ago, its most massive stellar components (e.g., $\ge$ 2 \msun, living $\leq$ 1.5 Ga) could have evolved onto the AGB on time to contribute stardust to the early Solar System. If this population, instead, migrated to the solar neighbourhood later on, it would have been less likely for it to reach this location in time to contribute stardust to the early Solar System, given that migration timescales are of the order of several Ga. 


The impact of the rates of the neutron source reactions, \iso{13}C($\alpha$,n)\iso{16}O and \iso{22}Ne($\alpha$,n)\iso{25}Mg recently and currently being measured in underground laboratories (LUNA and JUNA) needs to be carefully analysed, especially in relation to the \iso{13}C ingestion events. We will also need to consider and test the impact of the neutron-capture cross sections involved in the production of isotopes of interest. For example, new cross sections for the Nd isotopes would be specifically useful, as some disagreement is present between different studies \citep{guber97,wisshak98}. Our result will also needs to be checked against all the signatures observed in bulk meteorites as proposed by \cite{ek20} for the many different available elements (e.g., Sr, Zr, Mo, Ru, Pd, Ba, Nd, Sm, Dy, Os, W), as well as in leachates and CAIs. 

In terms of further improvements in the AGB models, we will need to calculate models with the new solar abundances \cite{magg22,lodders25}, with metallicity 0.0187, accounting for possible different compositions \citep{gustafsson25}, and for the composition of high-metallicity objects in the galactic solar neighbourhood \citep{nissen20}. Moreover, we will need to consider models of mass lower than those presented here, given that, for example, \cite{osborn25} required a minimum mass for carbon stars around 1.2 \msun\ to match the observed carbon star luminosity function in the Galaxy. 



To check for the self-consistency of the proposed scenario, a detailed comparison of the models presented here with the stardust grain data should also be carried out. In this respect, more data on the Nd isotopic composition in stardust grains, especially single grains, would help to better assess the source of the Nd variability. Also the $s$-process composition of graphite grains should be considered on top of that of SiC \cite{croat05,stephan26,pal26}, noting that these grains are more likely to form when the C abundance is higher, i.e., at lower metallicities \citep{ventura12b} and during the very final stages of the AGB phase.
Moreover, we need to check for the consistency and potential impact of the results presented here on the origin in the early Solar System of the short-lived radioactive nuclei \iso{107}Pd, \iso{135}Cs, \iso{182}Hf, and \iso{205}Pb \citep{trueman22,Leckenby24} produced by the $s$ process in AGB stars. Specifically, we need to analyse the effect of introducing a stardust component from a high-metallicity AGB population contributing to such isotopes, and specifically to the \iso{107}Pd/\iso{182}Hf ratio, which is another diagnostic of the first-to-second $s$-process peak relative abundances.  

From the experimental point of view, as discussed, the slope of $\sim$10 used here is a first estimate, which needs to be refined. For example,  
variations for Zr and Nd need to be measured in the same meteoritic samples for a large number of carbonaceous chondrites (e.g., from the CV, CM, CO, CK, CR groups) to be able to 
to better define the trends between Zr and Nd compared to using group averages. Precisely defining other meteoritic components, like CAIs, which carry strong nucleosynthetic variations, will also help to account for their effect on the variations in bulk chondrites. 

Finally, we remind that our aim was to test the hypothesis that the observed Zr and Nd relative variability represents an $s$-process signature. This is a first, most obvious hypothesis, because other heavy elements, such as Mo, Ru, W, and Hf also show an $s$-process signature \cite[e.g.,][]{worsham19IronRuMoW}. Moreover, the stardust SiC grains that carry such an $s$-process signature are present in the meteoritic inventory. As part of testing this first hypothesis, we have assumed here that these existing SiC grains carried the observed signature. The possible relevance of other neutron-capture processes \citep[i.e., the $i$, $n$, and $r$ processes, see][]{REVLugaro23} and other types of grains \citep{bizzarro25ice} will need to be investigated in the future.


%
%


\begin{acknowledgements}
This paper is based on work carried out as part of Work Package 9 of the ChETEC$-$INFRA project funded by the European Union’s Horizon 2020 research and innovation programme (ChETEC-INFRA -- Project no. 101008324), and as part of the Lend\"ulet Program LP2023-10 of the Hungarian Academy of Sciences.
ML and BSz thanks Evelin B\'anyai for help with python. We are all deeply grateful for the opportunity to contribute to this volume in memory of Roberto Gallino, who has been an invaluable colleague, teacher, mentor, supervisor, and friend. Roberto was one of the first researchers in stellar astrophysics able to understand and appreciate the link between models and bulk meteoritic data. We strived to write this paper following Roberto's instruction ``to be guided by the observations''. We thank the anonymous referee and Andrew Davis for their careful reading and commenting of the manuscript, which resulted in significant improvements. Footnote 3 was modified for clarification following a question from Victor Zhang from the Reviewer3 project.

\end{acknowledgements}


%
%



\end{document}